\documentclass[11pt,reqno]{article}
\usepackage{amsmath}
\usepackage{amsfonts}
\usepackage{amssymb}
\usepackage{hyperref}
\usepackage{latexsym}
\usepackage[dvips]{graphicx}
\usepackage{epsf}
\usepackage{color}

\allowdisplaybreaks \numberwithin{equation}{section}

\newcommand{\be}{\begin{equation}}
\newcommand{\ee}{\end{equation}}
\newcommand{\bea}{\begin{eqnarray}}
\newcommand{\eea}{\end{eqnarray}}

\newcommand{\f}{\frac}
\newcommand{\p}{\partial}
\newcommand{\na}{\nabla}
\newcommand{\Tr}{{\rm Tr}}

\let\a=\alpha  \let\b=\beta  \let\g=\gamma  \let\d=\delta
     \let\th=\theta   \let\l=\lambda
\let\m=\mu    \let\n=\nu          \let\r=\rho \let\om=\omega
\let\s=\sigma \let\t=\tau   \let\vph=\varphi   
    
\let\G=\Gamma    \let\L=\Lambda 
         
\let\Om=\Omega

\newcommand{\Vb}{\bar{V}}
\newcommand{\phib}{\bar{\phi}}

\newcommand{\at}{\tilde{a}}

\newcommand{\Ut}{\tilde{U}}
\newcommand{\Vt}{\tilde{V}}

\newcommand{\phit}{\tilde{\phi}}

\newcommand{\tD}{\tilde{\Delta}}

\newcommand{\cM}{\mathcal{M}}

\newcommand{\cP}{\mathcal{P}}
\newcommand{\cQ}{\mathcal{Q}}
\newcommand{\cR}{\mathcal{R}}

\begin{document}

\thispagestyle{empty}
\begin{flushright} \small
AEI-2014-008
\end{flushright}
\bigskip

\begin{center}
  {\LARGE\bfseries     Critical behavior in spherical and hyperbolic spaces}\\
[10mm]
{\large Dario Benedetti}
\\[3mm]
{\small\slshape
Max Planck Institute for Gravitational Physics (Albert Einstein Institute), \\
Am M\"{u}hlenberg 1, D-14476 Golm, Germany \\ 
\vspace{.3cm}
 {\upshape\ttfamily dario.benedetti@aei.mpg.de} 
} 
\end{center}
\vspace{5mm}

\hrule\bigskip

\centerline{\bfseries Abstract} \medskip

We study the effects of curved background geometries on the critical behavior of scalar field theory.
In particular we concentrate on two maximally symmetric spaces: $d$-dimensional spheres and hyperboloids.
In the first part of the paper, by applying the Ginzburg criterion, we find that for large correlation length the Gaussian approximation is valid on the hyperboloid for any dimension $d\geq 2$, while it is not trustable on the sphere for any dimension. This is understood in terms of various notions of effective dimension, such as the spectral and Hausdorff dimension.
In the second part of the paper, we apply functional renormalization group methods to develop a different perspective on such phenomena, and to deduce them from a renormalization group analysis. By making use of the local potential approximation, we discuss the consequences of having a fixed scale in the renormalization group equations. In particular, we show that in the case of spheres there is no true phase transition, as symmetry restoration always occurs at large scales. In the case of hyperboloids, the phase transition is still present, but as the only true fixed point is the Gaussian one, mean field exponents are valid also in dimensions lower than four.

\bigskip
\hrule\bigskip
\tableofcontents

\section{Introduction}
\label{Sec:intro}

Curved spaces in physics are typically associated to the setting of general relativity and cosmology. However, their relevance is of course much more general, and they appear for example in the classical mechanics of constrained systems, as well as in the study of membranes and interfaces in condensed matter. A brief review of theoretical and experimental motivations for studying the effects of curved geometry in condensed matter can be found in \cite{Sausset:2010}. The search and study of condensed matter systems characterized by an actual or effective curved geometry is also stimulated by the idea of analogue gravity \cite{Barcelo:2005fc}.

In this paper we will be interested primarily in the case in which the geometry is non-dynamical. Such situation is commonly considered in the cosmological setting as a first approximation in which the gravitational degrees of freedom are frozen, and one studies just a quantum field theory in curved spacetime. 
From the condensed matter perspective this can also be seen as a first approximation, or alternatively as the primary case of interest in situations where the curvature is introduced as a technical device (e.g. \cite{Caillol:1991}) or for theoretical modeling (e.g. \cite{Rubinstein:1983}).

The presence of curvature in the background geometry can have drastic effects on the infrared behavior of a model \cite{Callan:1989em}, and in particular on its phase transitions and critical behavior.
Much work has gone in this direction for the case of constant negative curvature, that is, for the case of statistical models in hyperbolic space. 
The differences between models in the usual flat background and in the hyperbolic one have been studied in the context of liquids \cite{Sausset:2008,Sausset:2009}, percolation \cite{Baek:2009perc,Sausset:2009perc}, Ising model \cite{Rietman:1992,Wu:1996,Doyon:2004fv,Shima:2006,Hasegawa:2007,Iharagi:2010}, XY model \cite{Baek:2009xy}, self-avoiding walks \cite{Madras:2005} and more.
Besides the hyperbolic case, it is worth mentioning also that curved spaces appear in the study of finite size effects \cite{Costa-Santos:2003a}, curvature defects \cite{Costa-Santos:2003b}, topological effects \cite{Hoelbling:1996} and of course in the presence of compactified dimensions \cite{Linhares:2012}.

Despite the many relevant works, many directions appear to be unexplored. In particular, a renormalization group approach to this kind of problems seems to be lacking. Of course the situation is quite different in the high-energy context, where renormalization group investigations on curved backgrounds are quite common. However, the focus there is typically on ultraviolet properties, at least until recently. Over the past few years there has been an increased interest on IR effects in the cosmological setting of de Sitter spacetime  (e.g. \cite{Burgess:2010dd,Garbrecht:2011gu,Serreau:2013psa,Akhmedov:2013vka}).
In particular, it has been noticed how nonperturbative renormalization group techniques can be applied to this context and it was showed that spontaneously broken symmetries are radiatively restored in de Sitter spacetime in any dimension \cite{Serreau:2013eoa}.
The de Sitter case, because of the Lorentzian signature of the metric, presents a number of technical challenges, and one would expect the situation to be somewhat easier in Euclidean signature. Surprisingly, as far as we know, there has not been a thorough study of this sort in Euclidean signature.

The purpose of this paper is to in part bridge such gap. We will study scalar field theory on two standard types of  curved Riemannian geometry, $d$-dimensional spheres and hyperboloids. Our goal will be to gain a detailed understanding of how the background curvature affects the critical behavior of the model at large distances. In Sec.~\ref{Sec:gauss} we will use the Ginzburg criterion in order to test when and whether we should expect that mean field gives trustable results. In this way we confirm general expectations based on effective dimensionality arguments, which we expand upon at the end of the section.
In the second part of the paper, Sec.~\ref{Sec:FRG}, we will explore more in detail the effects brought in by the curvature of space, making use of functional renormalization group techniques in the local potential approximation. We will in particular study the question of symmetry restoration (or existence of a phase transition), and more in general we will discuss how the presence of a dimensional external scale affects the usual renormalization group picture.
In order to keep the treatment as self-contained as possible, we include four appendices detailing the geometry of spheres and hyperboloids (App.~\ref{App:geom}), the spectra of their respective Laplace-Beltrami operators (App.~\ref{App:spectra}), the associated heat kernels (App.~\ref{App:HK}) and propagators (App.~\ref{App:prop}).

\section{The Gaussian approximation and effective dimensionality}
\label{Sec:gauss}

Let $(\cM,g_{\m\n})$ be a $d$-dimensional Riemannian manifold, that is, a differentiable manifold  $\cM$ equipped with a positive-definite metric $g_{\m\n}$ (in a given coordinate basis $x^\m$, $\m=1,\ldots,d$). The metric can be defined by the associated line element,\footnote{We use the Einstein convention, according to which repeated indices imply a summation.}
\be
ds^2_{(\cM)} = g_{\m\n} dx^\m dx^\n \, .
\ee
In this work we will restrict to homogenous spaces, and in particular we will consider only three types of $d$-dimensional Riemannian spaces: the flat space (as a benchmark), the sphere and the hyperboloid. The respective geometries are briefly reviewed in App.~\ref{App:geom}.

We are interested on the statistical properties of scalar field theories on such backgrounds.
The statistical field theory of the field $\vph=\vph(x)$ is characterized as usual by the generating functional
\be \label{genFunct}
Z[J] \equiv e^{W[J]} = \int D\vph \, e^{ -S[\vph] + \int d^dx \sqrt{g} J \vph } \, ,
\ee
and by the bare action
\be \label{bareS}
S[\vph] = \int d^d x \sqrt{g} \left[ \f{Z}{2} g^{\m\n} \p_\m \vph \p_\n \vph + U(\vph) \right] \, ,
\ee
where $g$ is the metric determinant, and $g^{\m\n}$ the inverse metric.
For the purpose of this Section, we choose
\be \label{bareV}
U(\vph)= \f12 (m^2+\xi R) \vph^2 + \f{u}{4!} \vph^4 \, ,
\ee
where $R$ is the Ricci scalar of the background, and $\xi$ a dimensionless coupling.
The mean-field approximation is obtained evaluating the partition function $Z[0]$ by saddle point method. For constant field the classical solution satisfies $U'(\vph_0)=0$, that is,
\be \label{orderparameter}
\vph_0 = \begin{cases} 0 &\,\, \mbox{ for }\,  m^2+\xi R>0\, , \\ \pm \sqrt{-6\f{m^2+\xi R}{u}} &\,\, \mbox{ for } \, m^2+\xi R<0\, .  \end{cases}
\ee
The transition from zero to non-zero mean field is a text-book example of second-order phase transition.

In the Gaussian approximation we keep also the quadratic fluctuations around the minimum of the potential, their covariance being given by the inverse of the second functional derivative (Hessian) of the action, evaluated at $\vph_0$: 
\be \label{Hessian}
S^{(2)} = \begin{cases} -Z\,\na^2 +m^2+\xi R &\,\, \mbox{ for }\,  m^2+\xi R>0\, , \\ -Z\,\na^2 - 2\, (m^2+\xi R) &\,\, \mbox{ for } \, m^2+\xi R<0\, .  \end{cases}
\ee
Here $\na^2$ is the Laplace-Beltrami operator (or simply the Laplacian) on the curved background, see \eqref{Lapl},
and given the structure of the Hessian we define the correlation length $\ell_c$,
\be \label{corrL}
\ell_c^{-2} \equiv b =  \begin{cases} \f{m^2+\xi R}{Z} &\,\, \mbox{ for }\,  m^2+\xi R>0\, , \\ - 2\f{m^2+\xi R}{Z} &\,\, \mbox{ for } \, m^2+\xi R<0\, .  \end{cases}
\ee
The correlation length \eqref{corrL} is the scale beyond which the flat-space propagator, i.e. the correlation between $\vph(x)$ and $\vph(x')$, decays exponentially with the distance $|x-x'|$, see \eqref{flat-asymp}. At the phase transition $\ell_c\to\infty$ and the propagator has a power-law behavior, see \eqref{flat-asymp0}.
On curved space, the definition of physical correlation length as scale at which exponential decay sets in requires some hierarchy of scales, as we have another length scale $a$ (see \eqref{R-Sd} and \eqref{R-Hd}). 
We will have a regime in which $a\gg \ell_c$, for which the same definition as on flat space applies, with the understanding that the exponential damping takes place for $a\gg\s(x,x')\gg\ell_c$, where $\s(x,x')$ is the geodesic distance between the two points. In the opposite regime, for $\ell_c\gg a$, the behavior is in general different than on flat space. For example, on the hyperboloid we find (using \eqref{G-hyper1-asymp}) that the propagator decays exponentially also for $\ell_c\to\infty$ as $\s(x,x')>a/(d-1)$, while on the sphere it makes no sense to talk of $\s(x,x')\gg a$ as $\s(x,x') \in [0, \pi a]$.  Nevertheless we will for convenience keep calling correlation length the one defined in \eqref{corrL}, as this acts as a control parameter for the transition in \eqref{orderparameter}, which classically is reached as $\ell_c\to\infty$.

A simple test for the validity of the Gaussian approximation for the description of the phase transition is given by the Ginzburg criterion (e.g. \cite{Kopietz:2010zz}), obtained by computing (in the broken phase) the quantity
\be \label{Qtest}
Q = \f{\int_{\ell_c} d^d x\sqrt{g} \, G(\s;\ell_c^{-2}) }{\int_{\ell_c} d^d x\sqrt{g}\, \vph_0^2 } \, ,
\ee
where the integrals extend over a region of radius $\ell_c$. Here $G(\s;\ell_c^{-2})$ is the correlation function, 
\be
G(\s;\ell_c^{-2}) = \f{ \d^2 W[J]}{\d J(x) \d J(x')}{}_{\Big{|}_{J=0}}\, ,
\ee
which on homogeneous spaces depends on the space points $x$ and $x'$ only via their geodesic distance $\s(x,x')$, and in the Gaussian approximation it is given by the inverse of \eqref{Hessian}. The correlation functions, or propagators, on curved backgrounds are reviewed in App.\ref{App:prop}.

If $Q\ll 1$, the fluctuations are small and the Gaussian approximation provides a good approximation. On the other hand, if $Q\gg 1$, fluctuations are large, the Gaussian approximation breaks down and we need a nonperturbative treatment.
At a second order phase transition, the correlation length diverges, hence we are interested in checking what happens to $Q$ in such limit.

In flat space, approximating the integral in the numerator with an integral over the whole space (exploiting the fact that the correlation function cuts off the integration at about a radius $\ell_c$, see \eqref{flat-asymp}), one finds
\be \label{Qflat}
Q \sim \f{\ell_c^{4-d} u}{3 Z^2} \f{d\, \pi\, \G(\f{d+1}{2}) }{(2\pi)^{\f{d+1}{2}} }\, ,
\ee
where we used $\vph_0^2 = 3 Z / (\ell_c^2 u)$. Note also that in the present section all propagators come with an extra factor $Z^{-1}$ with respect to the formulas reported in the appendix due to the difference between \eqref{Hessian} and the operator being inverted in \eqref{propEq}.
From \eqref{Qflat}, in the large-$\ell_c$ limit, we deduce the well-known critical dimension $d_c=4$, below which the Gaussian approximation does not provide a valid description of the phase transition.

\subsection{The hyperboloid}
\label{Sec:gaussH}

For the case of an hyperboloid $H^d$, we can use the results reported in the appendices. From App.~\ref{App:geom} we find the volume integral
\be
\int_{\ell_c} d^d x\sqrt{g} =a^d\, \Om_{d-1} \int_0^{\ell_c/a} dy \sinh(y)^{d-1} \sim a^d\, \Om_{d-1}  \f{e^{(d-1)\,\ell_c/a}}{(d-1)2^{d-1}}\, ,
\ee
where the last expression is obtained for $\ell_c/a\gg 1$.

As we did for the flat case, we evaluate the numerator  of \eqref{Qtest} by integrating over the whole space. Using the asymptotic expansion of the propagator \eqref{G-hyper1-asymp}
 for  $y=\s/a\gg 1$, which is analytical at $b=0$ (see also definitions \eqref{rho}, \eqref{om_pm}, \eqref{alpha_pm} and \eqref{beta_pm}), we find
\be
Q_{(H^d)}\sim \f{ e^{(\r-\om_+) \ell_c/a} \om_+^\r  }{\om_+ -\r  }  \f{2\r}{ e^{2\r\,\ell_c/a}} \f{u \ell_c^2 a^{2-d}}{3 Z^2}\, ,
\ee
and since $\om_+\to\r +\f{1}{2\r} \f{a^2}{\ell_c^2}$ for $\ell_c\to\infty$, we obtain
\be \label{Qhyper}
Q_{(H^d)} \sim  4 \r^{\r+2}  e^{-2\r\,\ell_c/a} \f{u \ell_c^4 }{3 Z^2 a^d}\to 0 \, , 
\ee
for any $d>1$ (i.e. $\r>0$). 
%
%
We conclude that on $H^d$ the Gaussian approximation provides a trustable description of the phase transition for any $d>1$.
Note that while the power law in \eqref{Qflat} could be deduced by simple dimensional analysis (being $\ell_c$ the only dimensionful parameter at play), in the case of curved background that is not the case anymore due to the presence of $a$, and an explicit calculation is needed to obtain the behavior in \eqref{Qhyper}.

\subsection{The sphere}
\label{Sec:gaussS}

The sphere is a compact space, and as we shall see, from this simple fact follow all its main distinctive features.
In fact, common wisdom would suggest that there should be no phase transition. However, the basic argument behind such expectation is based on lattice models, which at finite volume contain only a finite number of degrees of freedom, and hence they cannot give rise to a non-analytic behavior of the free energy or of any other thermodynamic function.
Here we are working in a continuous framework, where effectively the lattice spacing has been taken to zero,\footnote{In fact it is well known that in the continuum limit the interacting theory exists only for $d<4$. However, this observation is not very relevant for the Gaussian approximation we are discussing in the present section, while in the following section we will mostly restrict to $d=3$ when doing explicit calculations.} and the number of degrees of freedom is hence infinite even at finite volume. From such a perspective we see no reason a priori to expect or less a phase transition.
Therefore the logic that we follow here is that of being agnostic about it: starting from the mean field intuition, telling us that $\ell_c\to\infty$ corresponds to a phase transition, we will try first to understand whether the Gaussian is trustable in such limit, and only later we will come back to the issue of whether a phase transition exists or not for the sphere.

We now wish to apply the Ginzburg criterion to the case of a sphere.
Being a compact space, we can perform the integrals in \eqref{Qtest} directly over the whole space. At finite $\ell_c$, the denominator is a obviously finite number. In the numerator, a divergence could appear only from a singularity of the propagator, but the latter is only singular at the origin (see App.~\ref{App:prop}), where however it is exactly balanced by the vanishing of the volume element (just like in the flat and hyperbolic case). Therefore $Q_{(S^d)}$, evaluated on the whole sphere at finite $\ell_c$, is finite. We want to estimate its magnitude in the limit $\ell_c\to\infty$.
In such limit, the dominant contribution comes from the presence of a zero mode, which exists (in the sense of being a normalizable eigenfunction of the Laplace operator) and is isolated (the spectrum is discrete) precisely because the sphere is a compact space.
From \eqref{smallb-Gsphere} we find that in the massless limit, the propagator is dominated by the zero mode contribution $G_{(S^d)}(y;\ell_c^{-2})\sim \ell_c^2$ (the remainder $G^{(0)}_{(S^d)}(y;\ell_c^{-2})$ gives a finite contribution to the numerator of $Q_{(S^d)}$ in the $\ell_c\to\infty$ limit), and as a consequence
\be \label{Qsphere}
Q_{(S^d)} \sim \f{\ell_c^4 a^{-d} }{3Z} u\, .
\ee
Comparing to \eqref{Qflat} we conclude that, for large $\ell_c$, the effective behavior on $S^d$ is that of a zero-dimensional space,
and in particular the Gaussian approximation is expected to be insufficient at large correlation length for any $d$.

\subsection{Interpretation in terms of effective dimension}
\label{Sec:gaussDim}

The conclusions we have reached with the Ginzburg criterion could have also been guessed by a heuristic argument in terms of effective dimensionality. We are going to illustrate such an argument for two different notions of effective dimension, that is, the spectral and the Hausdorff dimension.
The former is defined as
\be \label{d_s}
d_s \equiv -2 \frac{\p \log \Tr [K(s)]}{\p \log s}\, ,
\ee
where $K(s)$ is the heat kernel for the Laplace-Beltrami operator (see App.~\ref{App:HK}, with $b=0$).
On flat space $d_s=d$, which justifies the definition, while on a general space it is in the limit of $s\to 0$ that we always have $d_s\to d$.
A simple interpretation of such property is that, $s$ being the diffusion time, small $s$ means that only a small neighborhood of a point is being explored by the diffusion process, hence the space looks flat at those scales.

For large $s$, curvature effects become important, and for $s\to+\infty$ (at $\ell_c^2 \gg s$) we find that $d_s\to 0$ on $S^d$, while  $d_s\to +\infty$ on $H^d$.
Such limits are easily found. For the sphere we use the spectral sum representation of the heat trace, which is convergent in the large-$s$ domain,
\be
\Tr [K_{(S^d)}(s)] = \f{1}{\Om_d a^d}   \sum_n D_n e^{-s \om_n} \, ,
\ee
from which we see that $\Tr [K_{(S^d)}(s)]\to \f{1}{\Om_d a^d}$ for $s\to+\infty$ (again due to the zero mode), and hence $d_s\to 0$.
This can be heuristically understood as the statement that the sphere looks like a point when observed from a very large distance.

For the hyperboloid, we can use \eqref{hypHK} to find that, for $d$ odd,
\be \label{trHK-Hd}
\Tr [K_{(H^d)}(s)] \propto \f{ e^{-s\r}  }{(4\pi s)^{d/2}} \, ,
\ee
where essentially the exponential decay is due to the presence of a ``mass gap'' in the spectrum and the absence of a zero mode. Plugging \eqref{trHK-Hd} into \eqref{d_s} we find $d_s=d + 2 s\r$, that is, the spectral dimension grows linearly with $s$. For $d$ even, the expression for the heat trace is complicated by the integral nature of the pseudo-differential operator, however 
using \eqref{hypHKeven} for $y=0$, we find that in the large $s$ limit the integral only contributes with subleading power corrections, and the dominant suppression is still given by the exponential factor in front, that for $b=0$ is again $e^{-s\r}$.
Therefore the same result is obtained for the spectral dimension at large $s$ in even as in odd dimensions.

We can also use a different notion of effective dimension, the Hausdorff dimension
\be
d_H = \f{\p \log \int_{L} d^d x\sqrt{g} }{\p \log L}\, ,  
\ee
where the integral extends over the set of points for which $\s(x,0)\leq L$.
For the sphere such integral reaches a plateau\footnote{Here and in all the rest of the paper the geodesic distance $\s(x,0)$ refers to the distance between $x$ and the origin along a direct geodesic path (i.e. one that does not wrap multiple times around the sphere). As on the sphere there are no points with $\s(x,0)> \pi a$, the volume of the ball stops increasing beyond $L=\pi a$.} at $L=\pi a$, hence the Hausdorff dimension is zero at large $L$.
On the contrary, for the hyperboloid the integral keeps growing exponentially, that is, faster than any power of $L$, and the Hausdorff dimension diverges.

It is well known that mean field theory becomes exact at large number of dimensions, hence the infinite effective dimensionality of the hyperboloid at large scales provides a heuristic explanation of the result we obtained from the Ginzburg criterion. At the same time, we know that in the Ising universality class there is no phase transition for $d\leq 1$, hence we might expect a failure of the Gaussian approximation for the sphere.

We should stress however that even though such arguments based on the effective dimension give a correct picture of the underlying physical mechanism, the Ginzburg criterion is more trustable as it involves directly the correlation function.

\section{A functional renormalization group perspective}
\label{Sec:FRG}

In this Section we want to analyze more in detail the effects induced by the curvature of the background geometry. To that end, we will use the method known as \emph{functional renormalization group} (FRG).\footnote{Sometimes referred to also as \emph{exact} or \emph{nonperturbative} renormalization group.}
There are many reviews on the FRG \cite{Morris:1998da,Bagnuls:2000ae,Berges:2000ew,Delamotte-review,Rosten:2010vm,Kopietz:2010zz}, to which we refer for an introduction to the topic.
We will employ here the FRG version that deals with the so-called effective average action \cite{Wetterich:1992yh}, 
\be \label{EAA}
\G_k[\phi] = \bar\G_k[\phi] -\f12 \int d^d x\sqrt{g}\, \phi\, \cR_k\, \phi
\ee
where $k$ stands for the running RG scale associated with the IR cutoff $\cR_k=\cR_k(-\na^2/k^2)$, and  $\bar\G_k[\phi]$ the Legendre transform of 
\be
W_k[J] = \log \int D\vph \, e^{ -S[\vph] + \int d^dx \sqrt{g} J \vph - \f12 \int d^d x\sqrt{g}\, \vph\, \cR_k\, \vph} \, ,
\ee
that is,
\be
 \bar\G_k[\phi] = \int d^d x\sqrt{g}\, J \phi - W_k[J]\, , \;\;\; \text{with} \;\; \phi = \f{\d W_k[J]}{\d J}\, ,
\ee
where in turn $W_k[J]$ is an IR-regulated version of the functional $W[J]$, introduced in \eqref{genFunct}. The effective average action satisfies the flow equation
\be\label{FRGE}
k\p_k \G_k = \f12 {\rm Tr} \left[ \left( \G_k^{(2)} 
+ \cR_k \right)^{-1} \, k\p_k  \cR_k  \right] \, ,
\ee
where $\G_k^{(2)}=\d^2\G_k/\d\phi\d\phi$ is the Hessian of $\G_k$. Note that while an additional UV cutoff $\L$ must be implicitly assumed in the definition of the functionals, the  cutoff $\cR_k$ is enough to render the equation \eqref{FRGE} finite both in the UV and in the IR, and therefore no reference to a UV cutoff is needed in \eqref{FRGE}. 
Similarly, in order to write down the equation no reference to the bare action is needed. However, in order to solve the flow equation we need initial conditions, and these will introduce the equivalent of a bare action and UV scale via $\G_{k=\L}[\phi]=S[\phi]$. Finally, from general properties of the IR cutoff (in particular $\lim_{k\to 0} \cR_k = 0$) it follows that upon integrating $\G_k[\phi]$ down to $k=0$ we obtain the full (textbook) effective action $\G[\phi]=\G_{k=0}[\phi]=\bar\G_{k=0}[\phi]$.

The equation \eqref{FRGE} is amenable to several approximation schemes, one of the most common being the derivative expansion, in which  $\G_k$ is expanded in invariants containing an increasing number of derivatives. 
The lowest order of the derivative expansion is known as local potential approximation (LPA), and it is the one we will consider here. 

The ansatz for the effective average action in the LPA is
\be \label{LPA}
\G_k[\phi] = \int d^d x \sqrt{g} \left[ \f{Z_k}{2} g^{\m\n} \p_\m \phi \p_\n \phi + V_k(\phi) \right] \, .
\ee
Strictly speaking, LPA usually stands for the case $Z_k=1$, otherwise \eqref{LPA} is typically referred to as LPA$'$ (e.g. \cite{Blaizot:2005wd}).
The functional RG equation \eqref{FRGE} for the potential reads
\be \label{kLPAeq}
k\p_k V_k(\phi) = \f12 \Tr_{(\cM)} \left[ \f{ k\p_k \cR_k(-\na^2/k^2)}{ - Z_k \na^2 + V''_k(\phi) +\cR_k(-\na^2/k^2)} \right]_{|_{\phi=\text{const.}} }\, ,
\ee
where we have redefined the trace on the Riemannian manifold $\cM$ by dividing out the volume of $\cM$, and we have projected the equation onto constant field configurations.

The equation \eqref{kLPAeq} describes the evolution of the potential under (continuous) coarse graining. The latter is the first of the two standard steps of the renormalization group \cite{Wilson:1973jj}, the second consisting in a rescaling of lengths and momenta such as to bring back the cutoff to its original value, and in a field redefinition that restores the normalization of the kinetical term.
In the LPA$'$, the second step is taken care of by the introduction of dimensionless variables
\be \label{dimless}
\phit = Z_k^{1/2}k^{(2-d)/2} \phi\, ,\;\;\; \Vt(\phit) = k^{-d} V(\phi(\phit)) \, .
\ee
We also define $\tD=-\na^2/k^2$, and we write the cutoff as  $\cR_k(z) = Z_k k^2 r(z)$, for some dimensionless function $r(z)$ constrained only by standard requirements  \cite{Litim:2001up}.
In dimensionless variables, \eqref{kLPAeq} reads
\be \label{LPAeq}
k\p_k \Vt_k(\phit) + d\, \Vt_k(\phit) - \f{d-2+\eta_k}{2} \phit \Vt'_k(\phit) = \widetilde{\Tr}_{(\cM)} \left[  \f{ (1-\eta_k/2) r(\tD) -\tD\, r'(\tD)   }{ \tD + \Vt''_k(\phit) + r(\tD)   }   \right]_{|_{\phit=\text{const.}} } \, ,
\ee
where $\eta_k= - k\p_k \ln Z_k$ is the scale-dependent anomalous dimension, and $\widetilde{\Tr}_{(\cM)} = k^{-d}\Tr_{(\cM)}$.
The LPA$'$ needs an additional equation for the flow of $Z_k$, which can be expressed as a relation between $\eta_k$ and (the derivatives of) the potential evaluated at its minimum (e.g. \cite{Berges:2000ew,Codello:2012sc,Wipf:2013vp}). In what follows, we will set $\eta_k=0$ in any practical calculation (i.e. we will only perform calculations within the strict LPA), so we will not need such expression. This is a common approximation in the flat case (see the already mentioned reviews  \cite{Morris:1998da,Bagnuls:2000ae,Berges:2000ew,Delamotte-review,Kopietz:2010zz} and references therein), where it is known to give qualitatively good results in general, as well as quantitatively accurate results in the case that the actual anomalous dimension is small (e.g. at the Wilson-Fisher fixed point in $d=3$, where $\eta\simeq 0.03$). We hope to come back to the LPA$'$ in future work, as this might be relevant for the $d=2$ case (see e.g. \cite{Codello:2012sc}).

A cutoff that leads to a very simple expression for the righ-hand-side of \eqref{kLPAeq} or  \eqref{LPAeq} is Litim's optimized cutoff \cite{Litim:2000ci,Litim:2001up}
\be \label{litim}
r(z)=(1-z)\th(1-z)\, ,
\ee
with which we obtain
\be \label{opteq}
k\p_k \Vt_k(\phit) + d\, \Vt_k(\phit) - \f{d-2+\eta_k}{2} \phit \Vt'_k(\phit) = \f{ 1  }{ 1 + \Vt''_k(\phit) } F_{(\cM)} (\at,\eta_k) \, ,
\ee
where
\be \label{optrhs}
F_{(\cM)}(\at,\eta_k)  =\widetilde{\Tr}_{(\cM)}[\th(1-\tD)] - \f{\eta_k}{2} \,\widetilde{\Tr}_{(\cM)}[(1-\tD)\th(1-\tD)] \, ,
\ee
and we have introduced
\be
\at=a k\, .
\ee

In order to explicitly perform the traces, we need to fix the dimension $d$. 
It is instructive to consider the case of $d=3$, for which computations are easiest, and where a nontrivial critical behavior is known to occur in the flat case.
In flat space, using Fourier transform we find
\be
F_{(E^3)}(\infty,\eta_k) = \f{\Om_{d-1} }{d\, (2\pi)^d}\left(1- \f{\eta_k}{d+2} \right)\Big{|}_{d=3} = \f{1}{6\pi^2} \left(1- \f{\eta_k}{5} \right)\, ,
\ee
and the analysis of the equation is standard (see e.g. \cite{Litim:2002cf,Bervillier:2007rc,Codello:2012sc,Wipf:2013vp}): one finds a non-trivial (Wilson-Fisher) fixed point, at which the critical exponents differ from their mean field value, and are in good agreement with the observed values.

On the hyperboloid, using the results collected in Appendix~\ref{App:spectra}, we find
\be \label{cfH3}
F_{(H^3)}(\at,\eta_k) =  \f{1}{6\pi^2}\left(1-\f{1}{a^2 k^2}\right)^{\f32} \th\left(1-\f{1}{a^2 k^2}\right) \left(1- \f{\eta_k}{5} \left(1-\f{1}{a^2 k^2}\right)\right) \, .
\ee
Finally, on the sphere we find
\be \label{cfS3}
F_{(S^3)}(\at,\eta_k) =\f{1}{a^3 k^3 \Om_3}\cP(\lfloor N_3 \rfloor ) \left( 1- \f{\eta_k}{2} \cQ(\lfloor N_3 \rfloor )  \right) \, ,
\ee
where   $\lfloor x \rfloor$ is the floor function,
\be
\cP(N)=\sum_{n=0}^{N} D_{n} = \f{1}{6} (1 + N) (2 + N) (3 +2 N) \, ,
\ee
\be
\cQ(N)= \f{1}{\cP(N)} \sum_{n=0}^{N} D_{n} (1-\tilde{\om}_n)=  \f{ 5 a^2 k^2 - 9 N -3 N^2  }{ 5 a^2 k^2}\, ,
\ee
being $\tilde\om_n$ the eigenvalues \eqref{eigS} in units of $k$, and
\be
N_3 =  -1 +  \sqrt{1+a^2 k^2 }  \, .
\ee
The spherical case gives rise to a staircase function, as a combined effect of the discrete spectrum and the use of a step function in the cutoff, a phenomenon already known in the literature (e.g. \cite{Benedetti:2012dx,Bilal:2013iva,Demmel:2014sga}).

We notice a crucial difference between the flat and the curved cases: in the curved backgrounds the FRG equation is a non-autonomous equation, in the sense that there is an explicit dependence upon $k$ on the rhs.
On flat space, it is the introduction of dimensionless variables that leads to an autonomous equation. In the curved background case, the existence of a fixed external scale implies that we cannot in general achieve an autonomous equation. The same thing generically happens if any non-running scale is present, for example in quantum field theory at finite temperature \cite{Berges:2000ew}, or on a non-commutative spacetime \cite{Gurau:2009ni}.

We thus immediately realize that true fixed points are unlikely, the potential will always retain a dependence on $k$ via its dimensionless product with $a$.
In special cases such dependence can be harmless, as in the case of the massless free theory. The latter is given by a $\phit$-independent potential $\Vt_k(\phit)=v_k$, with $v_k$ satisfying ($\eta_k=0$)
\be \label{vEq}
k\p_k v_k + d v_k  = \widetilde{\Tr}_{(\cM)} \left[  \f{ r(\tD) -\tD\, r'(\tD)   }{ \tD + r(\tD)   }   \right] \, .
\ee
Note that also on flat space the Gaussian solution to \eqref{opteq} has a non-zero vacuum term, $\Vt(\phit)=1/(d\,6\,\pi^2)$.
We could eliminate such running vacuum terms, and obtain a proper Gaussian fixed point with $\Vt_k(\phit)=0$, by a modified equation in which vacuum contributions are appropriately subtracted (see for example \cite{Kopietz:2010zz} or \cite{Gies:2006wv}).

Alternatively, we can introduce the concept of {\it floating-points} \cite{Gurau:2009ni}, i.e. solutions of the FRG equation which are independent of $k$, up to dependence on $\at=ak$.
In other words, we can introduce, and keep track of, an explicit dependence on $\at$, as if it was another field, an external field. Clearly, in the present case such procedure can be seen as a first step towards the treatment of cases in which the geometry is dynamical, and the curvature is indeed treated as on a par with other fields. We will discuss such point of view in Sec.~\ref{Sec:float}.

\subsection{Scaling dimension and symmetry restoration}
\label{Sec:scaling}

We will first discuss the consequences of the non-autonomy of the equation, taking the explicit formulas for $d=3$ with optimized cutoff as a guidance.

On the hyperboloid we observe that the loop contributions on the rhs of the FRG equation (i.e. the functional trace \eqref{cfH3}), vanish as soon as $k<1/a$, thus leaving us with the classical (tree level) part of the equation. Although we have not computed explicitly the functional traces needed to evaluate the anomalous dimension, it is clear that a similar phenomenon occurs also in such traces as all the functional traces always include the same step function coming from the cutoff, and hence the anomalous dimension also vanishes in the deep IR.
As a consequence, IR fixed points coincide with classical scale invariant theories, and thus mean field behavior is recovered at large distances, confirming our conclusions from Sec.~\ref{Sec:gaussH}.
It should be stressed that the use of a cutoff with step function, such as \eqref{litim}, provides us with an extreme version of the general case: with a generic cutoff the approach to zero will be smooth, but in general fast enough for $k<1/a$, thus leading to the same conclusion.

On the sphere, we see that $\cP(\lfloor N_3 \rfloor )\to 1$ for $k\to 0$, or more precisely as soon as $k^2<3/a^2$, that is, only the zero mode remains unsuppressed. However, the dimensionless volume of the 3-sphere goes to zero, making the rhs of the FRG equation divergent. 
A similar behavior happens in any dimension, as for  $k^2<d/a^2$ only the zero mode survives in the functional trace, which then reduces to the inverse of the volume (in units of $k$), i.e. $F_{(S^d)}(\at,0)=(\at^d \Om_d)^{-1}$ for $\at^2<d$.
The presence of a singularity for $k\to 0$ is a general consequence of the presence of compact dimensions, with the $d$-sphere behaving as $k^{-d}$ because all its dimensions are compact.
In order to absorb such divergence we should rescale the potential and the field such that the lhs be as divergent as the rhs. 
This is achieved by introducing the new variables (here and in the following $Z_k=1$)
\be \label{barphi}
\phib = (ka)^{d/2} \phit = a^{d/2} k \phi \, ,
\ee
\be \label{barpot}
\Vb(\phib) = (ka)^d \Vt(k^{-d/2} a^{-d/2} \phib ) = a^d V( a^{-d/2} k^{-1} \phib)\, .
\ee
The scaling of $\phib$ with $k$ has been chosen so that $1 + \Vt''_k(\phit)\to 1 + \Vb''_k(\phib)$.
The resulting equation (obtained by plugging (\ref{barphi}-\ref{barpot}) into \eqref{opteq}) for $k^2<d/a^2$ is
\be
k\p_k \Vb_k(\phib)  + \phib \Vb'_k(\phib) = \f{1}{ \Om_d}\, \f{ 1  }{ 1 + \Vb''_k(\phib) }\, ,
\ee
which can be recognized as the flat FRG equation for $d=0$, apart from the $\Om_d$ factor which could anyway be removed with a $k$-independent rescaling of field and potential.
We thus expect that the IR properties of scalar field theory on a spherical background will resemble those of the same theory in zero dimensions, in agreement also with the result \eqref{Qsphere}. In particular, as in flat space we have no phase transitions for $d<2$, we might expect this to be the fate also of scalar field theory on a sphere in any dimension.

Such expectation can be directly tested by studying the flow of the dimensionful potential, i.e. integrating \eqref{kLPAeq}, and looking for a transition in the IR between a potential with spontaneous symmetry breaking and one without.
What determines the presence or less of spontaneous symmetry breaking is the full effective potential, and this is obtained integrating the FRG equation all the way to $k=0$, starting with some initial condition at a UV scale $k=\L$. Such initial condition plays the same role that the bare action has in a path integral, and as such it is used to parametrize the phase diagram.
On flat space, this is a standard analysis (see for example \cite{Berges:2000ew,Kopietz:2010zz,Wipf:2013vp}), and it proceeds as following: one solves numerically the flow equation for the dimensionful potential with an initial condition at $k=\L$ corresponding to a potential with spontaneous symmetry breaking (potentials with minimum at the origin always remain such for the simple scalar field theory), i.e. $V_{\L}(\phi) = \l_\L (\phi^2- \r_\L)^2$ with $\r_\L>0$. Integrating down towards $k=0$ one observes in general that the local maximum at $\phi=0$ flattens out, as expected, because the full effective potential has to be convex (e.g. \cite{Haymaker:1983xk,O'Raifeartaigh:1986hi,Alexandre:2012hn}).\footnote{Note that the effective average action is not the Legendre transform of $W_k[J]$, as it differs from $\bar\G_k[\phi]$ by a cutoff term (see \eqref{EAA}). As a consequence, $\G_k[\phi]$ is not in general a convex functional. However, $\bar\G_k[\phi]$ should be convex as usual, hence we should have $\G_k^{(2)}+\cR_k\geq 0$ at any scale $k$, a property that we expect to be respected by the flow if we are careful in choosing initial conditions that respect such a constraint, and given that $\lim_{k\to 0} \cR_k = 0$, we should recover a convex effective action at $k=0$. The recovery of a convex $\G[\phi]=\G_{k=0}[\phi]$ has been studied in detail in the literature, see \cite{Berges:2000ew} and references therein.}
In this process, two different behaviors can arise depending on the initial condition $\r_\L$, namely in one case the potential becomes flat in a finite interval around the origin at $k=0$, corresponding to the effective potential of a broken phase,\footnote{The value at the left-most or right-most end of the flat region is chosen dynamically as the ground state when we switch off the external magnetic field (when the latter is on, the bottom of the potential is not horizontal, it has a slope, hence one of the two extremes of the linear region is a global minimum).} while in the other case the lowering of the maximum continues until we obtain at finite $k=k_s>0$ a global minimum at $\phi=0$, i.e. we obtain a symmetry restoration.
\begin{figure}[ht]
\begin{center}
\includegraphics[width=7.5cm]{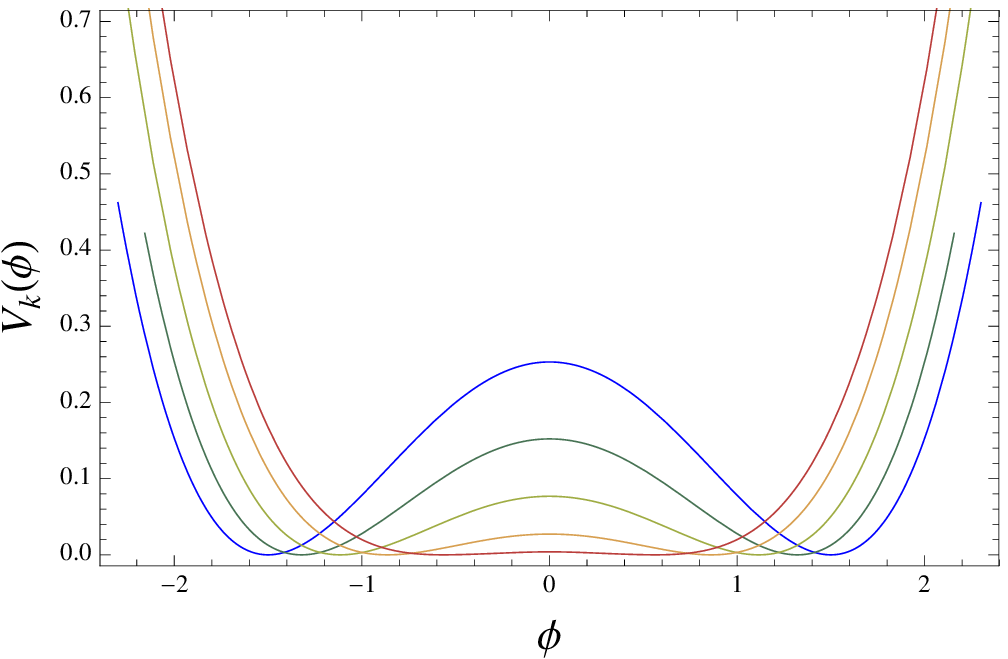} \hspace{.5cm}
\includegraphics[width=7.5cm]{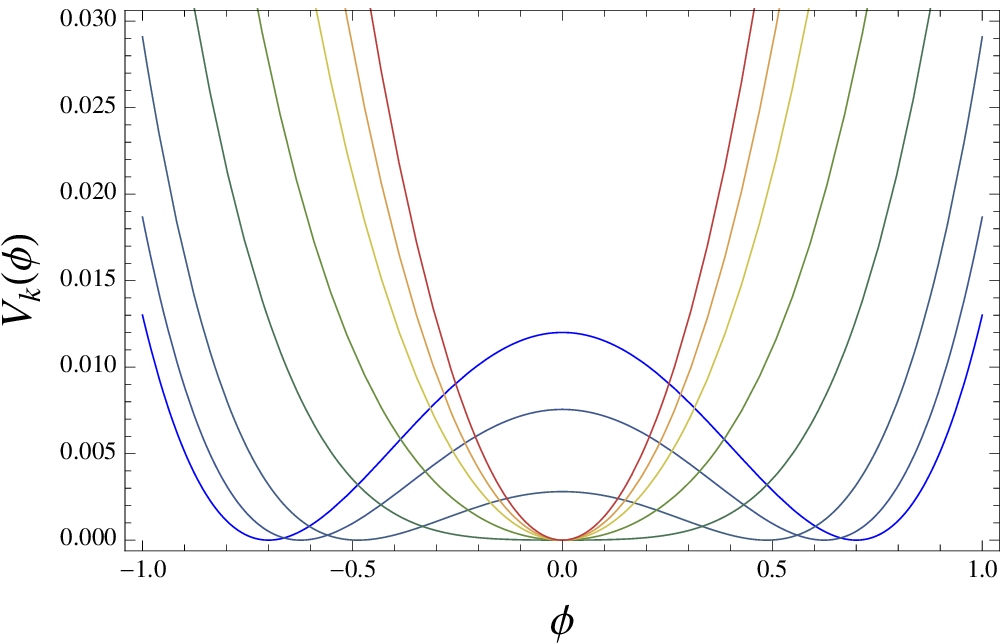}
\caption{
The flow of the potential (with the minimum subtracted for graphical purposes) on a flat background in $d=3$. The blue curve is the initial condition $V_{\L}(\phi) = \l_\L (\phi^2- \r_\L)^2$, with $\L=20$, $\l_\L=.05$, $\r_\L= 2.25$ (left) and $\r_\L=0.49$ (right). The red curve is at $k=0.1$, smaller values of $k$ being indistinguishable on the scale of the plot. Symmetry restoration is evident in the plot on the right. The phase transition occurs near $\r_\L \simeq 1.82$.}
\label{Fig:flat}
\end{center}
\end{figure}
On a flat background we know that both phases are present for $d\geq 2$ (or $d>2$ in the $O(N)$ model with $N>1$ \cite{Berges:2000ew,Codello:2012ec}), and a continuous phase transition separates them at some critical value $\r_\L=\r_c>0$. The behaviors characteristic of the two phases are depicted in Fig.~\ref{Fig:flat}.

\begin{figure}[ht]
\begin{center}
\includegraphics[width=7.5cm]{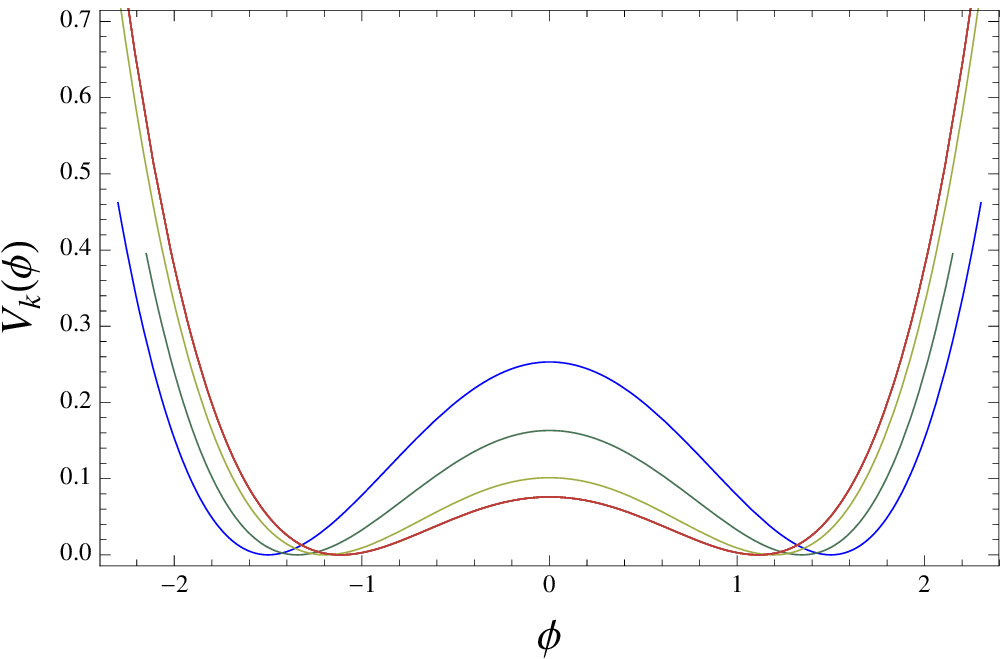} \hspace{.5cm}
\includegraphics[width=7.5cm]{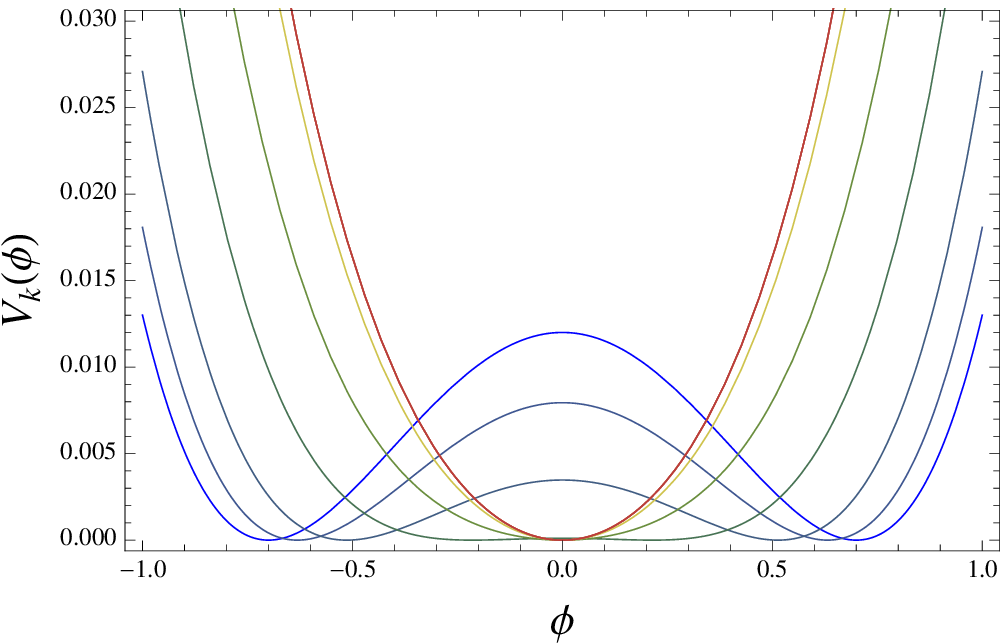}
\caption{
The flow of the potential (with the minimum subtracted) on a hyperbolic background in $d=3$. All the parameters and initial conditions are as in Fig.~\ref{Fig:flat}, and in addition we have  $a=1/5$. Although it looks like less curves are being plotted, the actual number is the same as in Fig.~\ref{Fig:flat}. However, as explained in the text, below a certain value of $k$ (for $k<5$ in this example) the potential freezes out and some of the curves are therefore superimposed.}
\label{Fig:hyper}
\end{center}
\end{figure}

On the hyperboloid, the phase diagram is qualitatively similar to the flat case, i.e. there exist both a broken and a symmetric phase, but the plots present an important difference, as can be seen in Fig.~\ref{Fig:hyper}.
The crucial point is that for $k<\r/a$ (compare \eqref{optrhs} and \eqref{eigH}) the rhs in \eqref{kLPAeq} vanishes, and the potential stops running. As a consequence, the potential remains frozen in the shape it had reached at $k=\r/a$, which could either be still a double well potential (left panel in Fig.~\ref{Fig:hyper}), or a symmetry-restored one (right panel in Fig.~\ref{Fig:hyper}), if symmetry restoration happens at $k=k_s\geq \r/a$. 
In the broken phase we seem to have a violation of the convexity property, but in fact there is no contradiction, as convexity of the effective action does not imply convexity of the effective potential on the hyperboloid.
Convexity of the effective action means that if we split the field as $\phi=\bar\phi+\vph$, we require $\vph\cdot\G^{(2)}[\bar\phi]\cdot\vph\equiv \int d^dx\sqrt{g(x)} \int d^dy\sqrt{g(y)} \vph(x)\G^{(2)}[\bar\phi](x,y)\vph(y)\geq 0$ for any $\vph$. Of course we can equivalently ask that all the eigenvalues of $\G^{(2)}[\bar\phi]$ are non-negative.
Working within a derivative expansion of the action, the potential acts multiplicatively on functions, and we only have to diagonalize the derivative terms. In flat space the diagonalization is implemented by plane waves, and $\G^{(2)}$ is reduced to a function of $\bar\phi$ and of the momentum $p_\m$. As the zero mode with $p^2=0$ is in the spectrum, it follows that the second derivative of the effective potential (i.e. $\G^{(2)}[\bar\phi]$ at $p^2=0$) has to be non-negative. On the contrary, in hyperbolic space the zero mode is not part of the spectrum, and this spoils the convexity argument for the effective potential. In fact, from \eqref{eigH} we know that the smallest eigenvalue of the Laplacian is $\n_0=\r^2/a^2$ (with eigenfunction $\vph_{0,l}$), and as a consequence  $\G^{(2)}[\bar\phi]\cdot\vph_{0,l} \neq V''(\bar\phi)\cdot\vph_{0,l}$.
For example, in the next to leading order approximation of the derivative expansion, where the constant $Z_k$ in \eqref{LPA} is replaced by a function $Z_k(\phi)$, for constant $\bar\phi$ we find $\G^{(2)}[\bar\phi]\cdot\vph_{0,l} = (Z(\bar\phi) \n_0 +V''(\bar\phi)) \vph_{0,l}$, showing that the potential itself (and in general any truncation of the derivative expansion) needs not be convex.
Interestingly this is agreement with the mean field approximation, in which the potential (as defined for example in \cite{Wipf:2013vp}) in the broken phase is not convex. We take this as another indication that on the hyperboloid mean field is valid at large scales.

\begin{figure}[ht]
\begin{center}
\includegraphics[width=8cm]{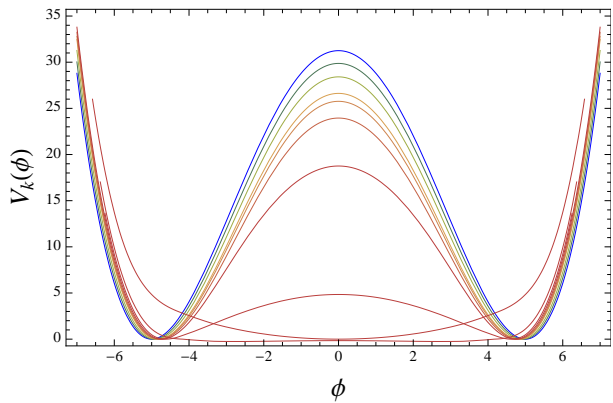} 
\caption{
The flow of the potential (again with the minimum subtracted) on a spherical background with $a=1/5$ in $d=3$. The blue curve is the initial condition $V_{\L}(\phi) = \l_\L (\phi^2- \r_\L)^2$, with $\L=20$, $\l_\L=.05$, $\r_\L= 25$, while the red curve is at $k=0.1$. Despite the large value of the initial symmetry breaking parameter, it is evident that symmetry restoration still takes place.}
\label{Fig:sphere}
\end{center}
\end{figure}

In the spherical case the situation is instead radically different, as it turns out that symmetry is always restored at finite $k$, i.e. we do not find the broken phase for any value of $\r_\L$. An example of symmetry restoration is shown in Fig.~\ref{Fig:sphere}. We find that, for large enough $\r_\L$, at some intermediate scale ($k\sim 0.6$ in the specific case of  Fig.~\ref{Fig:sphere}) the potential is basically that of a broken phase, but as we keep decreasing the scale the symmetry is restored by the development of a minimum at $\phi=0$.

\subsection{Floating points}
\label{Sec:float}

We can introduce an explicit dependence on $\at=a k$ in the potential, so as to transform the flow equation into an autonomous equation, but with an additional independent variable.
In order to highlight the presence of an additional argument in the potential, we denote the potentail as $U_k(\phi,a)$.
We then introduce the dimensionless potential
\be
\Ut_k(\phit,\at) = k^{-d} \, U_k(\phi(\phit),\at/k) \, ,
\ee
for which we obtain the equation
\be \label{float}
k\p_k \Ut_k(\phit,\at) + d\, \Ut_k(\phit,\at) - \f{d-2+\eta_k}{2} \phit\, \p_{\phit} \Ut_k(\phit,\at) + \at\, \p_{\at} \Ut_k(\phit,\at)
 = \f{ 1  }{ 1 +  \p^2_{\phit} \Ut_k(\phit,\at) } F_{(\cM)}(\at,\eta_k)  \, .
\ee
This is simply a rewriting of \eqref{opteq}, with $k\p_k \Vt_k(\phit) = k\p_k \Ut_k(\phit,\at) + \at\, \p_{\at} \Ut_k(\phit,\at)$.
Now the fixed point (or floating point) equation is provided by the PDE obtained by setting $k\p_k \Ut_k(\phit,\at)=0$.

In order to understand the meaning of \eqref{float} we can first consider the case in which we discard the loop contribution on the rhs. That is, we study the tree level equation (with $\eta_k=0$)
\be \label{float-tree}
d\, \Ut_k(\phit,\at) - \f{d-2}{2} \phit \, \p_{\phit} \Ut_k(\phit,\at) + \at\, \p_{\at} \Ut_k(\phit,\at)= 0\, .
\ee
Such equation, in the absence of boundary conditions simply constraints the dependence on the variables to
\be
\Ut_k(\phit,\at) = \at^{-d}\, Y\left(\at^{\f{d-2}{2}}\, \phit \right)\, ,\;\;\; \text{or}\;\;\;  U(\phi,a) = a^{-d} \, Y\left(a^{\f{d-2}{2}} \, \phi \right)\, ,
\ee
or in other words, the floating point potential is effectively a fixed point potential, where dimensional quantities (the potential and the field) are expressed in units of $a$.
This was to be expected as in the absence of the rhs, the FRG equation is simply a statement of classical scale invariance (or $k$-independence), and thus dimensional analysis is enough to fix the potential.
If in addition we require analyticity in both $\phi$ and $a^{-1}$ (the latter in order to recover the flat space limit), that is, if we require regular behavior at $a^{-1}=\phi=0$, and also $\mathbb{Z}_2$ symmetry, we find
\be
U(\phi,a) = \sum_{n=0}^{\lfloor \f{d}{d-2} \rfloor} c_n\, a^{-d+ n (d-2)}\, \phi^{2 n} \, ,
\ee
with free dimensionless coefficients $c_n$.
Because of the presence of the dimensionful parameter $a$, we find a more general potential than the usual $\phi^{\f{2d}{d-2}}$ required by scale invariance in flat space.

As another consequence of the dimensional scale given by the curvature, we can straightforwardly see that the Gaussian fixed point has a massive generalization that would be forbidden on flat space:
we can solve the full floating point equation with an ansatz of the type
\be
\Ut(\phit,\at) = u(\at) + c \, \f{\phit^2}{\at^2} \, .
\ee
When plugged into \eqref{float}, the second term disappears from the linear (or tree level) part of the equation, because it is a solution of \eqref{float-tree}. On the other hand, the trace part becomes $\phit$-independent, with $\p^2_{\phit} \Ut_k(\phit,\at)= 2 c/\at^2$. We are then left with an inhomogeneous linear ODE for $u(\at)$, for which the solution of the associated homogeneous equation is $u_{\rm hom}(\at)=u_0/\at^d$, for an arbitrary constant $u_0$, while the special solution of the inhomogeneous equation is scheme and dimension dependent.

Non-trivial floating points are much harder to study without truncations, as they require the flat case solution as boundary condition at $\at^{-1}=0$. 
For this reason, we will now resort to a polynomial truncation.

\subsection{A simple truncation}
\label{Sec:trunc}

Truncations of the potential to polynomial form are a very useful approximation even on flat space. The lowest order truncations can serve as a playground to understand qualitative features of the theory under examination (e.g. \cite{Bagnuls:2000ae,Berges:2000ew}), while their recursive extension can serve even as a quantitative method for the extraction of precise critical exponents (e.g. \cite{Canet:2003qd,Bervillier:2007rc}).
Here, since we focus on the qualitative picture rather than on precise quantitative estimates, we will consider the simplest possible truncation, that is, a simple quartic potential, and again $Z_k=1$.
We distinguish two cases, corresponding to equation \eqref{opteq} and \eqref{float} respectively:
\be \label{trunc1}
\Vt_k(\phit) = v_0(k) + v_2(k)\, \phit^2 + v_4(k)\, \phit^4 \, ,
\ee
and
\be \label{trunc2}
\Ut_k(\phit,\at) = u_0(k,\at) + u_2(k,\at)\, \phit^2 + u_4(k,\at)\, \phit^4 \, .
\ee

As observed previously, the first case \eqref{trunc1} does not lead to nontrivial fixed points.
We obtain the system of beta functions
\be \label{beta0}
k\p_k v_0 = -d \, v_0 + \f{F_{(\cM)}(\at,0)}{1+2 \, v_2} \, ,
\ee
\be \label{beta2}
k\p_k v_2 = -2 \, v_2 -12\, v_4\,  \f{F_{(\cM)}(\at,0)}{(1+2 \, v_2)^2} \, ,
\ee
\be \label{beta4}
k\p_k v_4 = (d-4) \, v_4 +144\, v_4^2\,  \f{F_{(\cM)}(\at,0)}{(1+2 \, v_2)^3} \, .
\ee
It is easy to check that setting the left-hand-sides to zero, if $F_{(\cM)}(\at,0)$ has a nontrivial dependence on $\at$ (i.e. in the non-flat case), then the only fixed point is at $v_2=v_4=0$ (as already discussed below \eqref{vEq}, in order to fix also $v_0=0$ we would need to modify the equation). The nontrivial solution is
\be \label{FP1}
v^*_2 = \f{4-d}{2 d - 32} \, , \;\;\; v^*_4 = \f{ 12 (d-4)}{(d-16)^3 F_{(\cM)}(\at,0)} \, .
\ee
In the flat case $F_{(\cM)}(\at,0)$ is a constant, and for $d<4$ this is the Wilson Fisher fixed point in the simplest truncation.
In the curved case this solution changes with $k$ (at fixed $a$), and we cannot interpret it as a fixed point. 
In fact, it is not a solution of the flow equation at all, because while it leads to the vanishing of the rhs of the flow equations, its lhs is non-vanishing. It could nevertheless be interpreted as fixed point in some limit ($k\to 0$ or $k\to\infty$) if the lhs vanishes in such limit too.
From the known behavior of  $F_{(\cM)}(\at,0)$, we find that as $k\to0$, $v^*_4\to +\infty$ in the hyperbolic case (actually as $k\to 1/a$ because of the optimized cutoff), while $v^*_4\to 0$ in the spherical case.
In the spherical case, the nontrivial solution merges with the Gaussian fixed point, but as we already know, there is no phase transition in this case.
In the hyperbolic case, the nontrivial fixed point is pushed to infinity, leaving us with only the Gaussian fixed point, thus explaining why the Gaussian approximation is valid in this case. 
We can also study the linear perturbations around \eqref{FP1} in $H^3$. As usual, we expand the beta functions \eqref{beta2} and \eqref{beta4} to linear order around the (would-be) fixed point \eqref{FP1}, and look for the eigenvalues of the matrix of the coefficients. The relevant eigenperturbations correspond to negative eigenvalues, and the latter are in close relation to the critical exponents (see for example \cite{Kopietz:2010zz}).
We find that the stability eigenvalues are $k$-independent and equal to $\n_{\pm} = \f16 (2 \pm \sqrt{82})$, but the eigendirections are $k$-dependent and become degenerate at $k=1/a$, both reducing to the vector $(0,1)$. As a consequence, stability eigenvalues have to be taken from the Gaussian fixed point, and thus they trivially coincide with the results from the Gaussian approximation (i.e. in $d=3$ they are $\n_2=-2$ and $\n_4=-1$).

In the parametrization \eqref{trunc2} we obtain
\be \label{beta0u}
k\p_k u_0 = - \at \p_{\at} u_0 -d \, u_0 + \f{F_{(\cM)}(\at,0)}{1+2 \, u_2} \, ,
\ee
\be \label{beta2u}
k\p_k u_2 = - \at \p_{\at} u_2 -2 \, u_2 -12\, u_4\,  \f{F_{(\cM)}(\at,0)}{(1+2 \, u_2)^2} \, ,
\ee
\be \label{beta4u}
k\p_k u_4 = - \at \p_{\at} u_4 + (d-4) \, u_4 +144\, u_4^2\,  \f{F_{(\cM)}(\at,0)}{(1+2 \, u_2)^3} \, .
\ee
Essentially what we have done here is to separate the $k$-dependence that is associated to the presence of the dimensionful parameter $a$, from any other $k$-dependence (see also below \eqref{float}). We now have a ``time''-independent system of partial differential equations, rather than a non-autonomous system of ordinary ones.
Imposing again the vanishing of  the left-hand-sides, we obtain this time a system of ordinary differential equations. The advantage of this formulation is that the solution to this ``floating point'' system of equations is now a true solution to the flow equations, differently from the status of  \eqref{FP1}.

The interesting case is the hyperboloid, which we can study once more in $d=3$. The equation for $\at>1$ is not easily integrated analytically but can of course be integrated numerically. However, whatever the solution is in that range, this has to be matched with the solution for $\at<1$. The latter is trivial because of the vanishing of $F_{(\cM)}(\at,0)$, and we are left with a set of equations that is essentially the same as the usual tree-level flow equations but with $k$ replaced by $\at$.
We thus obtain
\be \label{FP2}
u^*_0 = \f{c_0}{\at^3} \, , \;\;\; u^*_2 = \f{c_1}{\at^2} \, , \;\;\; u^*_4 = \f{c_2}{\at} \, ,
\ee
corresponding to the (dimensionful) potential
\be \label{U_FP}
U^*(\phi,a) = \f{c_0}{a^3} + \f{c_1}{a^2} \phi^2 + \f{c_2}{a} \phi^4 \, .
\ee
From \eqref{FP2} we obtain the same result as from \eqref{FP1}, i.e. that the dimensionless couplings (in units of $k$) go to infinity as $\at\to 0$. However, we stress again that \eqref{FP2} is an exact solution (for $k<a^{-1}$) of the system (\ref{beta0u}-\ref{beta4u}), while  \eqref{FP1} is not a solution of (\ref{beta0}-\ref{beta4}).
Furthermore,  \eqref{U_FP} gives a different and more interesting point of view on what is going on: due to the dimensionful scale $a$, we obtain a mean field $k$-independent potential, in agreement with Fig.~\ref{Fig:hyper}.

\section{Conclusions}
\label{Sec:concl}

In this paper we have studied the effects of curvature on the critical behavior of a scalar field, concentrating on spherical and hyperbolic spaces. 
By applying the Ginzburg criterion we have deduced that on a $d$-dimensional sphere the Gaussian approximation is never trustable when the correlation length becomes large, while on a $d$-dimensional hyperboloid it  is trustable for any $d \geq 2$. We have interpreted this in terms of effective dimensions, such as the spectral and Hausdorff dimension: in the far IR both notions of dimension indicate that spheres are effectively zero-dimensional (they look like a point) while hyperboloids have an infinite effective dimension. In view of the known dependence of the Ising universality class on the dimension, one would then expect to find no phase transition on the sphere, and to find a phase transition well described by mean field theory on the hyperboloid. Such expectations were confirmed in Sec.~\ref{Sec:FRG}, where we applied functional renormalization group techniques to the analysis of the scalar model on curved backgrounds. 
After discussing the general new features of the FRG equation in the presence of an external scale, we have shown by numerical integrations in the local potential approximation that on the sphere there is only the symmetric phase. Finally, with the help of a simple truncation, we have shown how the Wilson-Fisher fixed point is pushed to infinity in $H^3$, thus leaving us with the sole Gaussian fixed point, with trivial critical exponents.

The main purpose of this paper was to show how the FRG can help us understanding the effects of geometry on critical phenomena. To that end, we studied the simplest model, and tried to keep things simple, but many other extension and applications are of course possible. On the technical level, it would be desirable to consider smooth cutoffs, thus avoiding the nonanalytic staircase effects encountered with \eqref{litim}, and to study the LPA$'$ more in detail, as well as the full next-to-leading order of the derivative expansion.
A natural and simple extension of this work would be to study the $O(N)$ model, something to which we hope to come back in the near future. It would also be interesting to study what happens on different spaces, and in particular whether some space can be found in which a nontrivial behavior persists at the phase transition, but with different exponents than in the flat case.

\subsection*{Acknowledgements}
%

I would like to thank Alessandro Codello for useful conversations in the early stages of this work.

\newpage
\appendix
\section{Geometry of backgrounds}
\label{App:geom}

The most trivial homogeneous space, which we consider in this work as reference case, is flat Euclidean space $E^d$, with metric element $ds_{(E^d)}^2= \d_{\m\n} dx^\m dx^\n$. The other two spaces we study here are $d$-dimensional spheres and hyperboloids.

The $d$-sphere can be defined in an intrinsic way as the quotient $S^d\simeq SO(d+1)/SO(d)$, or in an extrinsic way via its embedding in $E^{d+1}$
\be
\sum_{A=1}^{d+1} (X^A)^2 = a^2\, ,
\ee
where $X^A$ are the Cartesian coordinates in $\mathbb{R}^{d+1}$, and $a$ is the radius of the sphere.
Its metric element can be written as
\be \label{dsS}
ds_{(S^d)}^2 = a^2 d\Om_d \equiv a^2 \sum_{i=1}^d  d\th_i^2 \prod_{j=i+1}^{d} \sin^2(\th_j) = a^2 d\th_d^2 + a^2 \sin^2(\th_d) d\Om_{d-1}\, ,
\ee
where the product is omitted for $i=d$.
The angles $\th_i$ take values in $[0,\pi]$, except for $\th_1\in[0,2\pi]$.
As any homogeneous space, the $d$-sphere is maximally symmetric, which implies that it is an Einstein space, i.e. $R_{\m\n}=\f1d g_{\m\n} R$ with constant scalar Ricci curvature R, and that it has zero Weyl tensor.
In other words, the Riemann tensor reduces to
\be
R_{\m\n\r\s} = \f{R}{d(d-1)} (g_{\m\r} g_{\n\s}-g_{\m\s}g_{\n\r})\, .
\ee
On $S^d$ the scalar Ricci curvature is given by
\be \label{R-Sd}
R_{(S^d)}=  \f{d(d-1)}{a^2} \, .
\ee
We often use the volume of the unit $d$-sphere, which is
\be \label{Om}
\Om_d \equiv a^{-d} \int_{(S^d)} d^d x \sqrt{g} = \f{ \G(d/2) }{ \G(d) } (4\pi)^{d/2} \, .
\ee

The $d$-dimensional hyperboloid is defined intrinsically as the quotient  $H^d\simeq SO(d,1)/SO(d)$, or extrinsically as the upper sheet ($X^{d+1}>0$) of the hypersurface
\be
\sum_{A=1}^{d} (X^A)^2 -(X^{d+1})^2 = - a^2\, ,
\ee
embedded in Minkowski space $M^{d,1}$, i.e. $\mathbb{R}^{d+1}$ with flat metric of signature $(+,\ldots,+,-)$.
Its metric element can be written as
\be \label{dsH}
ds_{(H^d)}^2 = d\t^2 + a^2 \sinh^2(\t/a) d\Om_{d-1} \, ,
\ee
where $d\Om_{d-1}$ is the metric element on the unit $(d-1)$-sphere, defined above, and $\t\in[0,+\infty)$ is the geodesic distance from the origin.
The dimensional parameter $a$ is the characteristic length or ``radius'' of the hyperboloid, in terms of which the scalar Ricci curvature is 
\be \label{R-Hd}
R_{(H^d)}= - \f{d(d-1)}{a^2} \, .
\ee
%

\section{Spectra of Laplacian operators}
\label{App:spectra}

On a generic Riemannian manifold with metric $g_{\m\n}$ the Laplace-Beltrami operator acting on a scalar field $\phi(x)$ is given by
\be \label{Lapl}
\na^2\, \phi(x) = \f{1}{\sqrt{g}} \p_\m (\sqrt{g} g^{\m\n} \p_\n \phi(x)) \, .
\ee

On flat space the eigenfunctions of the Laplacian are of course the plane waves, and the functional traces are evaluated via Fourier transform.
On curved backgrounds we lack a Fourier transform, however we are on a comparable situation whenever we know the spectrum of the Laplacian, as in the case of the spaces we consider in this work.

The Laplacian spectrum on the sphere is well known \cite{Rubin:1983be}, the scalar eigenmodes satisfying
\be \label{eigS}
-\na^2\, \psi_{n,j} = \frac{n(n+d-1)}{a^2} \,  \psi_{n,j} \equiv\om_n \,  \psi_{n,j} \, ,
\ee
with multiplicity $D_n=\frac{(n+d-2)!\, (2n+d-1)}{n!(d-1)!}$, $j=1,2,...D_n$, and $n=0,1,2,...+\infty$. 
Eingenmodes (whose explicit expression we do not need here) are orthonormal, that is,
\be \label{sphereNorm}
\int_{S^d} d^d x\sqrt{g}\, \psi_{m,j}^*(x) \psi_{m',j'}(x) = \d_{mm'} \d_{jj'} \, .
\ee

For the scalar Laplacian on the hyperboloid we follow \cite{Camporesi:1990wm,Camporesi:1991nw,Camporesi:1994ga}.
The eigenmodes of the Laplacian on $H^d$ satisfy
\be \label{eigH}
-\na^2 \phi_{\l,l} = \f{1}{a^2} (\l^2 + \r^2) \vph_{\l,l} \equiv\n_\l \,  \vph_{\l,l}\, ,
\ee
where 
\be \label{rho}
\r = (d-1)/2 \, ,
\ee
$\l\in[0,+\infty)$, and $l=0,1,2,...+\infty$.  
Eigenmodes are normalized as
\be
\int_{H^d} d^d x\sqrt{g}\, \vph_{\l,l}^*(x) \vph_{\l',l'}(x) = \d_{ll'}\, \d(\l-\l')  \, .
\ee
The analogue of the multiplicty for the continuum spectrum is the spectral function, or Plancherel measure, which is defined by
\be
\m(\l) \equiv \f{\pi \Om_{d-1} a^d}{2^{d-2} } \sum_l \vph_{\l,l}^*(0) \vph_{\l,l}(0) \, ,
\ee
and explicitly given by
\be
\m(\l) = \f{\pi}{ 2^{2(d-2)} \G(d/2)^2  } \prod_{j=0}^{(d-3)/2} (\l^2+j^2) \, 
\ee
for odd $d\geq 3$, and by
\be
\m(\l) = \f{\pi \l \tanh(\pi \l)}{ 2^{2(d-2)} \G(d/2)^2  } \prod_{j=1/2}^{(d-3)/2} (\l^2+j^2) \, 
\ee
for even $d\geq 2$ (for $d=2$ the product is omitted).

Functional traces (which we define divided by the volume) reduce to
\be \label{traceS}
\Tr_{(S^d)} [W(-\na^2)] =  \f{1}{\Om_d a^d}  \sum_n D_n W(\om_n)
\ee
for the sphere, and to
\be \label{traceH}
\Tr_{(H^d)} [W(-\na^2)] = \f{2^{d-2} }{\pi \Om_{d-1} a^d} \int_0^\infty d\l\, \m(\l) W(\n_\l) 
\ee
for the hyperboloid.

\section{Heat kernel}
\label{App:HK}

By definition the heat kernel is the solution of the heat equation
\be
(\p_s -\na_x^2 + b) K(x,s;x_0,b)=0 \, ,
\ee
with initial condition
\be
\lim_{s\to 0}K(x,s;x_0,b)= \f{\d(x-x_0)}{\sqrt{g}} \, .
\ee
On a homogeneous space the heat kernel depends only on the geodesic distance between $x$ and $x_0$, which we denote by $\s(x,x_0)$. 
We thus write $K(x,x_0,s;b)=K(\s,s;b)$.

Knowing the spectrum of the Laplacian we can write the general solution in the form
\be \label{genHK}
K(\s,s;b) = \sum_u e^{-s (\l_u+b)} \chi_u(x) \chi^*_u(x_0)\, ,
\ee
where $-\na^2 \chi_u(x) = \l_u \chi_u(x)$, and $u$ labels the whole set of eigenmodes.

On flat space we have
\be \label{flatHK}
K_{(E^d)}(x,s;b) = \f{ e^{-\f{|x|^2}{4s}-s b}  }{(4\pi s)^{\f{d}{2}}}\, .
\ee

For spheres and hyperboloids, the nearest we can get to a closed expression for the heat kernel on these spaces is probably in terms of fractional derivatives \cite{Camporesi:1990wm}. We introduce the dimensionless variable $y=\s/a$,  rescale $s\to a^2 s$, and define
\be \label{om_pm}
\om_\pm = \sqrt{\r^2 \pm  a^2 b} \, .
\ee
On the sphere one finds
\be \label{sphereHK}
K_{(S^d)}(y,s;b) = \f{1}{a^d} \f{ e^{s\om_-^2}  }{(4\pi s)^{\f12}} \left(\f{1}{2\pi} \f{\p}{\p (\cos(y)+1)} \right)^{\f{d-1}{2}} \sum_{n=-\infty}^{+\infty} (\pm 1)^n e^{-\f{(y+2\pi n)^2}{4s}}\, ,
\ee
where the plus and minus sign are for $d$ odd and even respectively.
As $y=\th_d$ (see \eqref{dsS}), we have that $y\in[0,\pi]$, however geodesics can wrap several times around the sphere, and such ``indirect paths'' precisely give rise to the sum over $n$ in \eqref{sphereHK}.

On the hyperboloid we have
\be \label{hypHK}
K_{(H^d)}(y,s;b) = \f{1}{a^d} \f{ e^{-s\om_+^2}  }{(4\pi s)^{\f12}} \left(-\f{1}{2\pi} \f{\p}{\p \cosh(y)} \right)^{\f{d-1}{2}} e^{-\f{y^2}{4s}}\, .
\ee
Note that the fractional derivatives have different definitions for the cases of the sphere and the hyperboloid \cite{Camporesi:1990wm}, but always reduce to ordinary derivatives for $d$ odd. Note also that the absence of indirect paths for the geodesics makes the expression for the hyperboloid simpler than that for the sphere.

Since in Sec.~\ref{Sec:gaussDim}  we need the behavior of \eqref{hypHK} for large $s$, we provide here some further detail on the fractional derivative for the hyperbolic case. The relevant definition is that of Weyl fractional derivative,
\be \label{fracDer}
\p^n_x f(x) = \f{(-1)^{-n}}{\G(-n)} \int_x^{+\infty} (x'-x)^{-n-1} \, f(x')\, dx'\, ,
\ee
which is well defined for any real $n<0$. For $n>0$ it is defined by analytic continuation, by using the property $\p^{p+q}_x f(x)=\p^p_x\p^q_x f(x)$, and choosing $p+q=n$, with $q=\lceil n \rceil$ and $p=n-q<0$.
We thus get that, for integer $m$,
\be \label{hypHKeven}
K_{(H^{2m})}(y,s;b) = \f{1}{a^d} \f{ e^{-s\om_+^2}  }{(4\pi s)^{\f12}}  \f{(-1)^{m} }{(2\pi)^{m-\f12}\G(\f12) } \int_y^{+\infty}  \, \left( \left(\f{\p^{m}}{\p \cosh(x') }\right)^{m} e^{-\f{x'^2}{4s}}\right)\,\f{ \sinh x' \, dx'}{(\cosh x'-\cosh y)^{\f12} } \, .
\ee
%

\section{Propagators}
\label{App:prop}

By definition the propagator $G(x,x_0;b)$ is the solution to the equation
\be \label{propEq}
(-\na_x^2 + b) G(x,x_0;b) = \f{\d(x-x_0)}{\sqrt{g}} \, .
\ee
Again, due to homogeneity of space the propagator depends only on $y=\s(x,x_0)/a$, hence we will simply write $G(y;b)$ for the propagator.
Its relation to the heat kernel is provided by the Schwinger proper time integral,
\be \label{G-K}
G(y;b) = a^2 \int_0^{\infty} ds K(y,s;b) \, ,
\ee
which, upon using \eqref{genHK}, gives (assuming that $\l_u+b>0$, $\forall u$)
\be \label{genG}
G(y;b) = \sum_u \f{1}{\l_u+b} \chi_u(x) \chi^*_u(x_0)\, .
\ee

On flat space the propagator is well known, and it takes the form (e.g. using \eqref{flatHK} and \eqref{G-K})
\be
G_{(E^d)}(x;b) = \f{b^{\f{d-2}{2}} }{(2 \pi )^{d/2} } (\sqrt{b}\, |x|)^{1-\frac{d}{2}} K_{\frac{d-2}{2}}\left(\sqrt{b}\, |x|\right) \, ,
\ee
where $K_{\n}(x)$ modified Bessel function of the second kind, leading to the asymptotic behavior
\be \label{flat-asymp}
G_{(E^d)}(x;b) \sim \f{b^{\f{d-2}{2}} }{(2 \pi )^{d/2} } (\sqrt{b}\, |x|)^{\frac{1-d}{2}} \sqrt{\f{\pi}{2}} \,e^{-\sqrt{b}\, |x|}
\ee
for $\sqrt{b}\, |x|\gg 1$, and
\be \label{flat-asymp0}
G_{(E^d)}(x;b) \sim \f{2^{\f{d-4}{2}} }{(2 \pi )^{d/2} } \G(\f{d-2}{2})\,  |x|^{2-d}
\ee
for  $\sqrt{b}\, |x|\ll 1$. This justifies the definition \eqref{corrL} of correlation length $\ell_c= b^{-1/2}$.

The propagators for both $S^d$ and $H^d$ have been computed in \cite{Allen:1985wd} directly solving \eqref{propEq}, or from an explicit mode sum in \cite{Dowker:1975tf} for the sphere and in \cite{Burgess:1984ti,Camporesi:1991nw} for the hyperboloid. Define
\be \label{alpha_pm}
\a_\pm = \r + \om_\pm\, ,
\ee
\be \label{beta_pm}
\b_\pm = \r -\om_\pm\, .
\ee

The propagator on $S^d$ is given by
\be \label{Gsphere}
G_{(S^d)}(y;b) = a^{2-d} \f{ \G(\a_-) \G(\b_-)}{\G(d/2)\, 2^d\, \pi^{d/2}}\, F(\a_-,\b_-;d/2; z) \, ,
\ee
where  $F(\a,\b;\g;z)$ is the hypergeometric function, and
\be
z = \cos^2(y/2)\, . 
\ee

Because of the existence of a constant mode $-\na^2\, \psi_{0,0}=0$, which by the normalization condition \eqref{sphereNorm} is $\psi_{0,0}(x)=a^{-d/2} \Om_d^{-1/2}$, the propagator on the sphere is singular at $b=0$. The singularity is easily isolated by writing (compare with \eqref{genG})
\be \label{smallb-Gsphere}
G_{(S^d)}(y;b) = \f{1}{b\, a^d\, \Om_d} + G^{(0)}_{(S^d)}(y;b) \, ,
\ee
where $G^{(0)}_{(S^d)}(y;b)$ is analytical in $b$ at $b=0$.

As in \eqref{sphereHK}, $y\in[0,\pi]$ and indirect geodesics are taken into account in the sum that leads to \eqref{Gsphere}. 
Furthermore, the propagator is regular at $y=\pi$, hence we only report on the small-$y$ (i.e. $z\sim1$) behavior, which reads
\be \label{Gsphere-sing}
G^{(0)}_{(S^d)}(y;b) \sim  \f{a^{2-d}}{2\,(4\pi)^{(d-1)/2}} \left( (1-z)^{1-\f{d}{2}}  f_1(z,b)+ f_2(z,b)+ f_3(z,b) \,\log(1-z)  \right)\, ,
\ee
where $f_i(z,b)$, $i=1,2,3$, are three analytical functions at $z=1$ and $b=0$, that depend on the dimension, and in particular $f_3(z,b) =0$ for odd dimensions. Their expression is not needed in this work hence we do not report it here.

The propagator on $H^d$ is given by
\be \label{G-hyper1}
G_{(H^d)}(y;b) = a^{2-d} \f{ \G(\a_+) \G(\a_+ -d/2+1)}{\G(\a_+ -\b_+ +1) 2^d \pi^{d/2}}  z^{-\a_+}\, F(\a_+,\a_+ -d/2+1;\a_+ -\b_+ +1; z^{-1}) \, ,
\ee
where
\be
z = \cosh^2(y/2)\, . 
\ee

The asymptotic behaviour for $y\gg 1$ reads
\be \label{G-hyper1-asymp}
G_{(H^d)}(y;b) \sim a^{2-d} \f{ \G(\a_+) \G(\a_+ -d/2+1)}{\G(\a_+ -\b_+ +1) 2^d \pi^{d/2} } \,4^{\a_+}\,  e^{-\a_+ y} \, ,
\ee
while for $y\ll 1$ we have
\be \label{Ghyper-sing}
G_{(H^d)}(y;b) \sim  \f{a^{2-d}}{2\,(4\pi)^{(d-1)/2}} \left( (1-z)^{1-\f{d}{2}} \tilde f_1(z,b)+\tilde f_2(z,b)+\tilde f_3(z,b) \,\log(1-z)  \right)\, ,
\ee
where again $\tilde f_i(z,b)$, $i=1,2,3$, are three analytical functions at $z=1$ and $b=0$, that depend on the dimension, and in particular $\tilde f_3(z,b) =0$ for odd dimensions.


\providecommand{\href}[2]{#2}\begingroup\raggedright\endgroup


\begin{thebibliography}{10}

\bibitem{Sausset:2010}
F.~Sausset, G.~Tarjus and P.~Viot, {\it {Statistical mechanics of liquids and
  fluids in curved space}},  Adv. Chem. Phys. {\bf 148} (2011) 251--310
  [\href{http://arXiv.org/abs/1005.2684}{{\tt arXiv:1005.2684}}].

\bibitem{Barcelo:2005fc}
C.~Barcelo, S.~Liberati and M.~Visser, {\it {Analogue gravity}},  Living
  Rev. Rel. {\bf 8} (2005) 12 [\href{http://arXiv.org/abs/gr-qc/0505065}{{\tt
  arXiv:gr-qc/0505065}}].

\bibitem{Caillol:1991}
J.~M. Caillol and D.~Levesque, {\it Numerical simulations of homogeneous and
  inhomogeneous ionic systems: An efficient alternative to the ewald method},
  J. Chem. Phys. {\bf 94} (1991) 597--607.

\bibitem{Rubinstein:1983}
M.~{Rubinstein} and D.~R. {Nelson}, {\it {Dense-packed arrays on surfaces of
  constant negative curvature}},  Phys. Rev. {\bf B28} (1983) 6377--6386.

\bibitem{Callan:1989em}
J.~Callan, Curtis~G. and F.~Wilczek, {\it {Infrared behavior at negative
  curvature}},  Nucl. Phys. {\bf B340} (1990) 366--386.

\bibitem{Sausset:2008}
F.~{Sausset}, G.~{Tarjus} and P.~{Viot}, {\it {Tuning the fragility of a
  glass-forming liquid by curving space}},  Phys. Rev. Lett. {\bf 101} (2008)
  155701 [\href{http://arXiv.org/abs/0805.1475}{{\tt arXiv:0805.1475}}].

\bibitem{Sausset:2009}
F.~{Sausset}, G.~{Tarjus} and P.~{Viot}, {\it {Thermodynamics and structure of
  simple liquids in the hyperbolic plane}},  J. Stat. Mech. {\bf 4} (2009) 22
  [\href{http://arXiv.org/abs/0903.0745}{{\tt arXiv:0903.0745}}].

\bibitem{Baek:2009perc}
S.~K. {Baek}, P.~{Minnhagen} and B.~J. {Kim}, {\it {Percolation on hyperbolic
  lattices}},  Phys. Rev. {\bf E79} (2009) 011124
  [\href{http://arXiv.org/abs/0901.0483}{{\tt arXiv:0901.0483}}].

\bibitem{Sausset:2009perc}
F.~{Sausset}, C.~{Toninelli}, G.~{Biroli} and G.~{Tarjus}, {\it {Bootstrap
  percolation and kinetically constrained models on hyperbolic lattices}},  J.
  Stat. Phys. {\bf 138} (2010) 411--430
  [\href{http://arXiv.org/abs/0907.0938}{{\tt arXiv:0907.0938}}].

\bibitem{Rietman:1992}
R.~Rietman, B.~Nienhuis and J.~Oitmaa, {\it The Ising model on hyperlattices},
  J. Phys. A: Math. Gen. {\bf 25} (1992) 6577.

\bibitem{Wu:1996}
C.~Wu, {\it Ising models on hyperbolic graphs},  J. Stat. Phys. {\bf 85} (1996)
  251--259.

\bibitem{Doyon:2004fv}
B.~Doyon and P.~Fonseca, {\it {Ising field theory on a pseudosphere}},
  J. Stat. Mech. {\bf 0407} (2004) P07002
  [\href{http://arXiv.org/abs/hep-th/0404136}{{\tt arXiv:hep-th/0404136}}].

\bibitem{Shima:2006}
H.~{Shima} and Y.~{Sakaniwa}, {\it {Geometric effects on critical behaviours of
  the Ising model}},  J. Phys. A: Math. Gen. {\bf 39} (2006) 4921--4933
  [\href{http://arXiv.org/abs/cond-mat/0511539}{{\tt arXiv:cond-mat/0511539}}].

\bibitem{Hasegawa:2007}
I.~{Hasegawa}, Y.~{Sakaniwa} and H.~{Shima}, {\it {Novel scaling behavior of
  the Ising model on curved surfaces}},  Surface Science {\bf 601} (2007)
  5232--5236 [\href{http://arXiv.org/abs/cond-mat/0612509}{{\tt
  arXiv:cond-mat/0612509}}].

\bibitem{Iharagi:2010}
T.~{Iharagi}, A.~{Gendiar}, H.~{Ueda} and T.~{Nishino}, {\it {Phase transition
  of the Ising model on a hyperbolic lattice}},  J. Phys. Soc. Japan {\bf 79}
  (2010) 104001 [\href{http://arXiv.org/abs/1005.3378}{{\tt arXiv:1005.3378}}].

\bibitem{Baek:2009xy}
S.~K. {Baek}, H.~{Shima} and B.~J. {Kim}, {\it {Curvature-induced frustration
  in the XY model on hyperbolic surfaces}},  Phys. Rev. {\bf E79} (2009) 060106
  [\href{http://arXiv.org/abs/0811.1895}{{\tt arXiv:0811.1895}}].

\bibitem{Madras:2005}
N.~Madras and C.~Wu, {\it Self-avoiding walks on hyperbolic graphs},  Comb.
  Prob. and Comp. {\bf 14} (2005) 523--548.

\bibitem{Costa-Santos:2003a}
R.~{Costa-Santos} and B.~M. {McCoy}, {\it {Finite size corrections for the
  Ising model on higher genus triangular lattices}},  J. Stat. Phys. {\bf 112}
  (2003) 889--920 [\href{http://arXiv.org/abs/cond-mat/0210059}{{\tt
  arXiv:cond-mat/0210059}}].

\bibitem{Costa-Santos:2003b}
R.~{Costa-Santos}, {\it {Logarithmic corrections to the Ising model free energy
  on lattices with conical singularities}},  Phys. Rev. {\bf B68} (2003) 224423
  [\href{http://arXiv.org/abs/cond-mat/0306396}{{\tt arXiv:cond-mat/0306396}}].

\bibitem{Hoelbling:1996}
C.~Hoelbling and C.~Lang, {\it {Universality of the Ising model on sphere-like lattices}},
  Phys. Rev. {\bf B54} (1996) 3434
  [\href{http://arXiv.org/abs/hep-lat/9602025}{{\tt arXiv:hep-lat/9602025}}].

\bibitem{Linhares:2012}
C.~Linhares, A.~Malbouisson and I.~Roditi, {\it GinzburgÐLandau theory of phase
  transitions in compactified spaces},  Advances in Quantum Field Theory, Prof.
  Sergey Ketov (Ed.), InTech (2012).

\bibitem{Burgess:2010dd}
C.~Burgess, R.~Holman, L.~Leblond and S.~Shandera, {\it {Breakdown of
  semiclassical methods in de Sitter space}},  JCAP {\bf 1010} (2010) 017
  [\href{http://arXiv.org/abs/1005.3551}{{\tt arXiv:1005.3551}}].

\bibitem{Garbrecht:2011gu}
B.~Garbrecht and G.~Rigopoulos, {\it {Self regulation of infrared correlations
  for massless scalar fields during Inflation}},  Phys. Rev. {\bf D84} (2011)
  063516 [\href{http://arXiv.org/abs/1105.0418}{{\tt arXiv:1105.0418}}].

\bibitem{Serreau:2013psa}
J.~Serreau and R.~Parentani, {\it {Nonperturbative resummation of de Sitter
  infrared logarithms in the large-N limit}},  Phys. Rev. {\bf D87} (2013)
  085012 [\href{http://arXiv.org/abs/1302.3262}{{\tt arXiv:1302.3262}}].

\bibitem{Akhmedov:2013vka}
E.~Akhmedov, {\it {Lecture notes on interacting quantum fields in de Sitter
  space}},  Int. J. Mod. Phys. {\bf D23} (2014), no.~1 1430001
  [\href{http://arXiv.org/abs/1309.2557}{{\tt arXiv:1309.2557}}].

\bibitem{Serreau:2013eoa}
J.~Serreau, {\it {Renormalization group flow and symmetry restoration in de
  Sitter space}},  Phys. Lett. {\bf B730} (2014) 271--274
  [\href{http://arXiv.org/abs/1306.3846}{{\tt arXiv:1306.3846}}].

\bibitem{Kopietz:2010zz}
P.~Kopietz, L.~Bartosch and F.~Schutz, {\it {Introduction to the functional
  renormalization group}},  Lect. Notes Phys. {\bf 798} (2010) 1--380.

\bibitem{Morris:1998da}
T.~R. Morris, {\it {Elements of the continuous renormalization group}},  Prog.
  Theor. Phys. Suppl. {\bf 131} (1998) 395--414
  [\href{http://arXiv.org/abs/hep-th/9802039}{{\tt arXiv:hep-th/9802039}}].

\bibitem{Bagnuls:2000ae}
C.~Bagnuls and C.~Bervillier, {\it {Exact renormalization group equations: An
  introductory review}},  Phys. Rept. {\bf 348} (2001) 91
  [\href{http://arXiv.org/abs/hep-th/0002034}{{\tt arXiv:hep-th/0002034}}].

\bibitem{Berges:2000ew}
J.~Berges, N.~Tetradis and C.~Wetterich, {\it {Non-perturbative renormalization
  flow in quantum field theory and statistical physics}},  Phys. Rept. {\bf
  363} (2002) 223--386 [\href{http://arXiv.org/abs/hep-ph/0005122}{{\tt
  arXiv:hep-ph/0005122}}].

\bibitem{Delamotte-review}
B.~Delamotte, {\it An introduction to the nonperturbative renormalization
  group},  in {\em Renormalization Group and Effective Field Theory Approaches
  to Many-Body Systems} (A.~Schwenk and J.~Polonyi, eds.), vol.~852 of {\em
  Lecture Notes in Physics}, pp.~49--132.
\newblock Springer Berlin Heidelberg, 2012.
\newblock \href{http://arXiv.org/abs/cond-mat/0702365}{{\tt
  arXiv:cond-mat/0702365}}.

\bibitem{Rosten:2010vm}
O.~J. Rosten, {\it {Fundamentals of the exact renormalization group}},
  Phys. Rept. {\bf 511} (2012) 177--272
  [\href{http://arXiv.org/abs/1003.1366}{{\tt arXiv:1003.1366}}].

\bibitem{Wetterich:1992yh}
C.~Wetterich, {\it {Exact evolution equation for the effective potential}},
  Phys. Lett. {\bf B301} (1993) 90--94.

\bibitem{Blaizot:2005wd}
J.-P. Blaizot, R.~Mendez-Galain and N.~Wschebor, {\it {Non perturbative
  renormalisation group and momentum dependence of n-point functions (I)}},
  Phys. Rev. {\bf E74} (2006) 051116
  [\href{http://arXiv.org/abs/hep-th/0512317}{{\tt arXiv:hep-th/0512317}}].

\bibitem{Wilson:1973jj}
K.~Wilson and J.~B. Kogut, {\it {The Renormalization group and the epsilon
  expansion}},  Phys. Rept. {\bf 12} (1974) 75--200.

\bibitem{Litim:2001up}
D.~F. Litim, {\it {Optimised renormalisation group flows}},  Phys. Rev. {\bf
  D64} (2001) 105007 [\href{http://arXiv.org/abs/hep-th/0103195}{{\tt
  arXiv:hep-th/0103195}}].

\bibitem{Codello:2012sc}
A.~Codello, {\it {Scaling solutions in continuous dimension}},  J. Phys. {\bf
  A45} (2012) 465006 [\href{http://arXiv.org/abs/1204.3877}{{\tt
  arXiv:1204.3877}}].

\bibitem{Wipf:2013vp}
A.~Wipf, {\it {Statistical approach to quantum field theory}},  Lect. Notes
  Phys. {\bf 864} (2013).

\bibitem{Litim:2000ci}
D.~F. Litim, {\it {Optimisation of the exact renormalisation group}},  Phys.
  Lett. {\bf B486} (2000) 92--99
  [\href{http://arXiv.org/abs/hep-th/0005245}{{\tt arXiv:hep-th/0005245}}].

\bibitem{Litim:2002cf}
D.~F. Litim, {\it {Critical exponents from optimized renormalization group
  flows}},  Nucl. Phys. {\bf B631} (2002) 128--158
  [\href{http://arXiv.org/abs/hep-th/0203006}{{\tt arXiv:hep-th/0203006}}].

\bibitem{Bervillier:2007rc}
C.~Bervillier, A.~Juttner and D.~F. Litim, {\it {High-accuracy scaling
  exponents in the local potential approximation}},  Nucl. Phys. {\bf B783}
  (2007) 213--226 [\href{http://arXiv.org/abs/hep-th/0701172}{{\tt
  arXiv:hep-th/0701172}}].

\bibitem{Benedetti:2012dx}
D.~Benedetti and F.~Caravelli, {\it {The local potential approximation in
  quantum gravity}},  JHEP {\bf 1206} (2012) 017
  [\href{http://arXiv.org/abs/1204.3541}{{\tt arXiv:1204.3541}}].

\bibitem{Bilal:2013iva}
A.~Bilal and F.~Ferrari, {\it {Multi-loop Zeta function regularization and
  spectral cutoff in curved spacetime}},  Nucl. Phys. {\bf B877} (2013)
  956--1027 [\href{http://arXiv.org/abs/1307.1689}{{\tt arXiv:1307.1689}}].

\bibitem{Demmel:2014sga}
M.~Demmel, F.~Saueressig and O.~Zanusso, {\it {RG flows of Quantum Einstein
  Gravity on maximally symmetric spaces}},
  \href{http://arXiv.org/abs/1401.5495}{{\tt arXiv:1401.5495}}.

\bibitem{Gurau:2009ni}
R.~Gurau and O.~J. Rosten, {\it {Wilsonian renormalization of noncommutative
  scalar field theory}},  JHEP {\bf 0907} (2009) 064
  [\href{http://arXiv.org/abs/0902.4888}{{\tt arXiv:0902.4888}}].

\bibitem{Gies:2006wv}
H.~Gies, {\it {Introduction to the functional RG and applications to gauge
  theories}},  Lect. Notes Phys. {\bf 852} (2012) 287--348
  [\href{http://arXiv.org/abs/hep-ph/0611146}{{\tt arXiv:hep-ph/0611146}}].

\bibitem{Haymaker:1983xk}
  R.~W.~Haymaker and J.~Perez-Mercader,
  {\it {Convexity of the Effective Potential}},
  Phys.\ Rev.\ D {\bf 27} (1983) 1948.

\bibitem{O'Raifeartaigh:1986hi}
  L.~O'Raifeartaigh, A.~Wipf and H.~Yoneyama,
  {\it {The Constraint Effective Potential}},
  Nucl.\ Phys.\ B {\bf 271} (1986) 653.

\bibitem{Alexandre:2012hn}
  J.~Alexandre,
  {\it {Spontaneous symmetry breaking and linear effective potentials}},
  Phys.\ Rev.\ D {\bf 86} (2012) 025028
  [\href{http://arXiv.org/abs/1205.1160}{{\tt arXiv:1205.1160}}].


\bibitem{Codello:2012ec}
A.~Codello and G.~D'Odorico, {\it {O(N)-Universality classes and the
  Mermin-Wagner theorem}},  Phys. Rev. Lett. {\bf 110} (2013), no.~14 141601
  [\href{http://arXiv.org/abs/1210.4037}{{\tt arXiv:1210.4037}}].

\bibitem{Canet:2003qd}
L.~Canet, B.~Delamotte, D.~Mouhanna and J.~Vidal, {\it {Nonperturbative
  renormalization group approach to the Ising model: A Derivative expansion at
  order partial**4}},  Phys. Rev. {\bf B68} (2003) 064421
  [\href{http://arXiv.org/abs/hep-th/0302227}{{\tt arXiv:hep-th/0302227}}].

\bibitem{Rubin:1983be}
M.~A. Rubin and C.~R. Ordonez, {\it {Eigenvalues and degeneracies for
  n-dimensional tensor spherical harmonics}},  J. Math. Phys. {\bf 25 (10)}
  (1975) 2888--2894.

\bibitem{Camporesi:1990wm}
R.~Camporesi, {\it {Harmonic analysis and propagators on homogeneous spaces}},
  Phys. Rept. {\bf 196} (1990) 1--134.

\bibitem{Camporesi:1991nw}
R.~Camporesi, {\it {Zeta function regularization of one loop effective
  potentials in anti-de Sitter space-time}},  Phys. Rev. {\bf D43} (1991)
  3958--3965.

\bibitem{Camporesi:1994ga}
R.~Camporesi and A.~Higuchi, {\it {Spectral functions and zeta functions in
  hyperbolic spaces}},  J. Math. Phys. {\bf 35} (1994) 4217--4246.

\bibitem{Allen:1985wd}
B.~Allen and T.~Jacobson, {\it {Vector two point functions in maximally
  symmetric spaces}},  Commun. Math. Phys. {\bf 103} (1986) 669.

\bibitem{Dowker:1975tf}
J.~Dowker and R.~Critchley, {\it {Effective Lagrangian and energy momentum
  tensor in de Sitter space}},  Phys. Rev. {\bf D13} (1976) 3224.

\bibitem{Burgess:1984ti}
C.~Burgess and C.~Lutken, {\it {Propagators and effective potentials in Anti-de
  Sitter space}},  Phys. Lett. {\bf B153} (1985) 137.

\end{thebibliography}
\end{document}